\documentclass[usenatbib]{mn2e}

\usepackage{graphics}
\usepackage{epsfig}
\usepackage{natbib}

\begin{document}

\title[Inner Rotation Curves]
{ A systematic study of the inner rotation curves of galaxies observed as part of the GASS and COLD GASS surveys} 

\author [G.Kauffmann et al.] {Guinevere Kauffmann$^1$\thanks{E-mail: gamk@mpa-garching.mpg.de},
 Mei-Ling Huang$^1$, Sean Moran$^2$, Timothy M.  Heckman$^3$\\  
$^1$Max-Planck Institut f\"{u}r Astrophysik, 85741 Garching, Germany\\
$^2$Harvard-Smithsonian Center for Astrophysics, 60 Garden Street,
 Cambridge, MA 02138, USA\\
$^3$Department of Physics \& Astronomy, Johns Hopkins University,
Baltimore, MD, 21218, USA}

\maketitle

%===================================
\begin{abstract} 
We present a systematic analysis of the rotation curves of 187 galaxies
with masses greater than $10^{10} M_{\odot}$,  with atomic  gas masses
from the GALEX Arecibo Sloan Survey (GASS), and with follow-up long-slit
spectroscopy from the MMT. Our analysis focuses
on stellar rotation curves derived by fitting stellar template spectra to
the galaxy spectra binned along the slit. In this way, we are able to obtain accurate rotation velocity
measurements for a factor of 2 more galaxies than possible
with the H$\alpha$ line.  Galaxies with high atomic gas
mass fractions are the most dark-matter dominated galaxies in our sample
and have dark matter halo density profiles that are well fit by Navarro,
Frenk \& White profiles with an average concentration parameter of 10.
The inner slopes and  of the rotation curves correlate  more strongly with
stellar population age than with galaxy mass or structural parameters. At
fixed stellar mass, the rotation curves of more actively star-forming galaxies have steeper inner
slopes  than less actively star-forming galaxies.  The ratio between the
galaxy specific angular momentum and the total specific angular momentum
of its dark matter halo, $R_j$, correlates strongly with galaxy mass,
structure and gas content. Low mass, disk-dominated galaxies with
atomic gas mass fractions greater than 20\% have median values of $R_j$
of around 1, but massive, bulge-dominated galaxies have $R_j=0.2-0.3$.
We argue that these trends can be understood in a picture where gas
inflows triggered by disk instabilities lead to the formation of passive,
bulge-dominated galaxies with low specific angular momentum.
\end{abstract}


\section {Introduction}

Rotation curves are an important probe of the distribution of both the
luminous and the dark matter mass in galaxies. They have been studied since the
1970's primarily as a way to constrain the density profiles of the dark
matter halos that are thought to surround all galaxies.  Because dark
matter increasingly dominates the mass distribution at large distances
from the centre of a galaxy, most observational studies have focused on
rotation curves measured using the HI line at 21 cm wavelengths. This
is because the atomic gas in galaxies generally extends significantly
further out than the stars. Many galaxies, however, have holes in the HI
distribution at small radii, so if one wishes to measure the shape of the
dark matter potential over a wide range of scales, the HI data needs to be
complemented by rotation measurements from optical H$\alpha$ or CO mm line
data (see Sofue \& Rubin 2001; Combes 2002 for comprehensive reviews).

There is a significantly smaller body of work on the statistical
properties of rotation curves.  Persic et al. (1996) compiled 1100
rotation curves of spiral galaxies and claimed that a single functional
form with galaxy luminosity as the only parameter could provide a
``universal'' description. The most luminous galaxies show  slightly
declining rotation curves in the outer part, following a maximum in the
disk. Intermediate galaxies have nearly flat rotation curves across
the disk. The least luminous galaxies have monotonically increasing
rotation curves across their whole disk. These results led Persic et al
(1996) to conclude that the dark-to-luminous mass ratio increases with
decreasing galaxy luminosity and mass.  Later studies of early-type disc
galaxies (S0-Sab) by Sofue et al (1999) and  Noordermeer et al (2006)
revealed large discrepancies with the rotation curves predicted by the
Persic et al. fitting formula. These authors found that the shape of the
rotation curve  depends on the bulge-to-disk ratio of the galaxy;
galaxies with concentrated light distributions and larger bulges reach
their maximum rotation velocities at smaller radii than galaxies with
smaller bulges and more diffuse light distributions.

One limitation of past statistical studies is that because they rely
on H$\alpha$, HI or CO line data, they are biased to galaxies with
high gas content and/or ongoing star formation. In principle, galaxy
rotation curves can also be derived from stellar absorption lines,
but this requires high signal-to-noise spectra and accurate fitting of
stellar templates in order to accurately measure shifts in velocity
along the slit. In recent years, this has become possible thanks to the
increasing efficiency of optical spectrographs and the availability
of high resolution stellar template spectra spanning a wide range in
wavelength (e.g. Martinsson et al 2013a).

Moran et al. (2012) presented long-slit spectroscopy of 174 galaxies drawn
from the GALEX Arecibo Sloan Digital Sky Survey (SDSS)(GASS; Catinella
et al. 2010), which was designed to measure the neutral hydrogen content
of a representative sample of $\sim 1000$ galaxies with stellar masses
larger than $10^{10} M_{\odot}$ and redshifts $0.02 < z < 0.05$ uniformly
selected from the SDSS (York et al. 2000) and Galaxy Evolution Explorer
(GALEX; Martin et al 2005) imaging surveys.  GASS observations detect HI
down to a gas fraction limit of $\sim$ 3\%. A companion project on the
IRAM 30m telescope, COLD GASS (Saintonge et al 2011), has obtained CO
(1-0) line measurements and  molecular gas masses for a subset of these
galaxies. In addition, a variety of stellar population and structural
parameters derived from GALEX/SDSS imaging and spectroscopy are available
for the full sample.

This paper presents a systematic analysis of the rotation curves derived
from the final sample of 236 longslit spectra obtained over the period
from October 2008 to April 2012 using both the Blue Channel Spectrograph
on the 6.5m MMT telescope and the Dual Imaging Spectrograph on the 3.5m
telescope at Apache Point Observatory (APO). As part of the procedure for
measuring SFR and gas-phase metallicity profiles for the galaxies in the
sample,  Moran et al (2012) derived both stellar rotation curves  from
fitting template stellar absorption line spectra to the galaxy spectra
binned along the slit, and H$\alpha$ rotation curves from fitting  Gaussian
profiles to the H$\alpha$ emission line after subtraction of the best-fit spectrum.

As we will show, the agreement between the H$\alpha$ and stellar rotation
measurements are generally good to within 10-20\% in galaxies where both
can be accurately determined. We are able to obtain accurate stellar
rotation velocity measurements for a factor of 2 more galaxies than is
possible with H$\alpha$ alone (187 out of 236 galaxies), permitting
a much wider exploration of galaxy parameter space.  The stellar
rotation curves generally do not extend beyond $\sim$2.5$R_e$ and thus
do not in general sample the flat part of the rotation curve with high
accuracy. Nevertheless, we will show that we can still use them as
probes of, a) the inner mass distribution of the galaxy, b) the radius
at which dark matter begins to dominate over baryons (stars and gas), c)
the angular momentum content of the galaxy as parametrized by the spin
parameter $\lambda$. After presenting our data analysis  methodology
in Section 2, we examine how these three quantities depend on a variety
of galaxy properties in Section 3. In Section 4 , we select a subset of
galaxies where dark matter dominates well within the optical radius, and
we examine constraints on the shape of the dark matter density profile. In
section 5, we summarize and conclude. We adopt a Hubble constant of $H_0=
70$ km s$^{-1}$ Mpc$^{-1}$ throughout the paper.

\section {Data Analysis Methodology}

\subsection {Rotation curve estimation and sample selection}

The procedure for measuring rotation curves is described in detail in Moran 
et al (2010). Working outwards from the galaxy centre, each  spectrum is binned
spatially to achieve a signal-to-noise of at least 6 (\AA $^{-1}$) in each
bin, the minimum needed for a reliable velocity estimate. For each binned
spectrum, the effective spatial position is determined by calculating the 
luminosity-weighted average radius of all spatial positions that entered the
co-add.

The radial velocity as a function of radius is measured in two ways: 1) by
cross-correlating each spectrum against superpositions of template spectra
from Bruzual \& Charlot (2003) stellar population synthesis models to 
determine the velocity, 2) by fitting directly for the centroids of the 
H$\alpha$ emission lines. The methods yield generally consistent curves at radii 
where we can fit templates and measure emission line centroids. This is shown
in Figure 1, where we show plots of the difference between the velocity determined by
fitting stellar templates  ($V_{abs}$) and the velocity infered from the  shift of the
centroid of the H$\alpha$ line ($V_{H\alpha}$). The velocity differences are
normalized by dividing by   $ 0.5(V_{abs} + V_{H\alpha})$ to yield a {\em fractional}
difference. We compute the average fractional distance for each galaxy
with where $V_{abs}$ and $V_{H\alpha}$
are both well-measured over at least 10 spectral bins, and we plot this quantity as  a function of  different global galaxy
parameters in each panel of Figure 1.

\begin{figure}
\includegraphics[width=80mm]{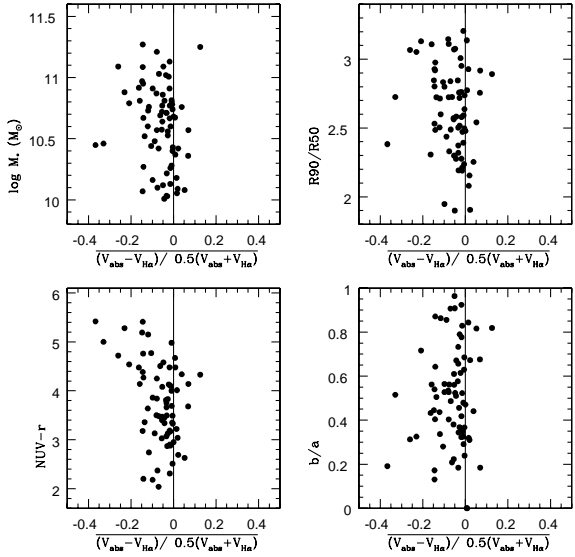}
\caption{ The mean fractional difference between the  velocity estimated by
fitting stellar templates to the longslit spectra ($V_{abs}$) and by fitting
to the centroid of the H$\alpha$ emission line ($V_{H\alpha}$) is plotted as a function 
of different global galaxy parameters in each panel: stellar mass
(top left), $r$-band concentration parameter (top right), global NUV-r
colour (bottom left), and axial ratio (bottom right).
\label{models}}
\end{figure}

As can be seen, $V_{H\alpha}$ is  generally larger
than $V_{abs}$, but this is a small effect (10\% systematic effect on average).   
A similar effect was found by Martinsson et al (2013a), which these authors attributed to
``asymmetric drift'', i.e. the fact that stars can be
supported against gravitational collapse by both rotation
and by random motions. If this is the case, one might expect the average velocity
difference to correlate strongly with the bulge-to-disk ratio of the galaxy, and for the
velocity difference to get systematically larger in the inner regions of the galaxy.
We can explore this hypothesis by plotting the velocity difference as a function of $\Sigma_*$
for each spectral bin, where $\Sigma_*$ is the local stellar surface mass density of
stars measured from our derived stellar mass profiles. This is shown in Figure 2.
Black points show the fractional difference between $V_{H\alpha}$ and  $V_{abs}$
for individual spectral bins, while the red squares show the average value of
the fractional offset in fixed bins of $\Sigma_*$. As can be seen, the average
fractional offset increases from 0.03 at stellar surface densities less
than $10^8 M_{\odot}$ kpc$^{-2}$ to 0.15 at stellar surface densities of 
$10^9 M_{\odot}$ kpc$^{-2}$, which is consistent with the ``asymmetric drift''
hypothesis. The bottom panels of Figure 1 show that the largest
velocity diferences averaged over the rotation curves
of individual galaxies  occur for galaxies with red NUV-r colours and high inclinations. 
This might be caused by dust extinction effects
or systematic errors in the measurment of the centroid of the  H$\alpha$
when the signal-to-noise is low. In either case we expect the stellar rotation measurements to be
more reliable.                                                      
In what follows, we will analyze the stellar rotation curves
without attempting to correct for asymmetric drift or extinction effects, which would
require more detailed modelling.   

\begin{figure}
\includegraphics[width=80mm]{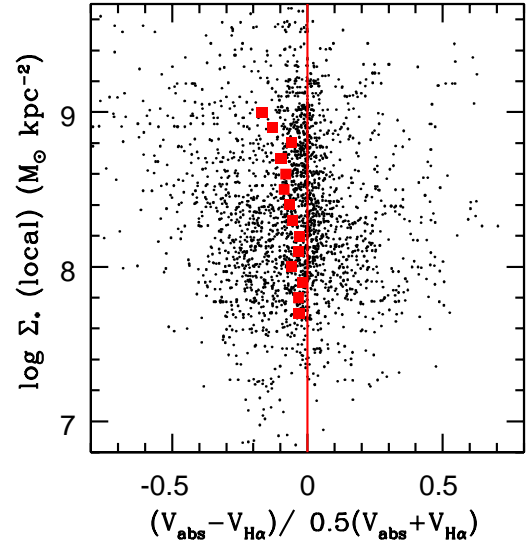}
\caption{ The fractional difference between the  velocity estimated by
fitting stellar templates to the longslit spectra ($V_{abs}$) and by fitting
to the centroid of the H$\alpha$ emission line ($V_{H\alpha}$) is plotted as a function 
of local stellar mass surface density $\Sigma_*$ Black points show results for individual
spectral bins, while red squares show the average value of the fractional difference
as a function of $\Sigma_*$.
\label{models}}
\end{figure}

We  proceed to inspect each of the 236 stellar rotation curves by eye and we select
those that have measured velocities with fractional errors less than 20\% across
a significant portion of the optical disk and those that are regular enough
to fit with a smooth function of the form
\begin{equation} 
V(R)= V_{max}R/(R^{\alpha} + r^{\alpha}_s)^{1/\alpha}+\Delta V
\end {equation}
(B\"ohm et al. 2004; Moran et al 2007,2010), where $R$ is the radius, $\alpha$ and $r_s$
are free parameters that govern the shape of the rotation curve, and $\Delta V$ is the offset of the
galaxy's central velocity from the redshift obtained from the SDSS spectrum (also
left as a free parameter). This results in a sample of 187 galaxies, of which 89 
(i.e. around a half) also have H$\alpha$ rotation curves measured over the
same radii as the  stellar rotation curves. The properties of this sample are shown
in Figure 3. The black histograms show the stellar mass, concentration index and NUV-r colour
distributions of the selected sample of 187 galaxies, while the red histograms are
for the galaxies where the rotation curves were either very poorly measured or too
irregular to work with. As can be seen, the selected sample of 187 galaxies with good
stellar rotation curves spans
a wide range in stellar mass, concentration and colour;  only the very most massive
($M_* > 10^{11} M_{\odot}$), reddest (NUV-r$>$ 6) and most concentrated ($R_{90}/R_{50}> 3.25$)
galaxies are systematically excluded.

\begin{figure}
\includegraphics[width=85mm]{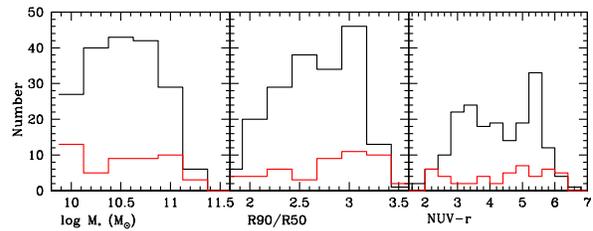}
\caption{ The black histograms show the distribution of the number of galaxies as a function
of stellar mass (left), concentration parameter (middle) and NUV-r colour (right)
for our selected sample of 187 galaxies with well-measured, regular stellar rotation curves.
The red histograms show the distributions for the 49 galaxies where stellar rotation curves
could not be accurately determined, or where the rotation curves were too irregular
to fit a function of the form given in the text.
\label{models}}
\end{figure}

\subsection {Parameters derived from the rotation curves}

Our stellar rotation curves are well-determined out to radii of
between 7 and 20 kpc. This is generally not far out enough in the disk to measure the shape 
of the outer rotation curve, but we are able to accurately measure the inner slope
and determine whether the gravitational potential is dominated by baryons (stars+gas)
or by dark matter over the optical radius of the disk. In addition, as we will
discuss, we  measure the total angular momentum content of the stellar disk, as 
parameterized by the spin parameter $\lambda$.

\begin{enumerate}
\item {\em The inner slope} is defined to be the slope of the best-fit
linear relation between $V_{abs}$ and $R$, evaluated from the center of the galaxy out to a radius of
$\pm r_s$, where $r_s$ is the fitting parameter in equation  (1). 
The inner slope provides a measure of the shear in the central regions of the galaxy.
In order to be able to compare massive galaxies with less massive galaxies in a meaningful
way, we scale the velocities by dividing by $V_{max}$, where $V_{max}$ is an estimate of the
maximum rotation velocity of the galaxy. We also scale the radii $R$  the radii by dividing by $R_{50}$, where
$R_{50}$ is the radius containing half the total r-band light of the galaxy.
Because the errors on the fitted value of $V_{max}$ in equation (1) are large for galaxies where
the outer regions of the rotation curve are not very well sampled, we   
estimate $V_{max}$ using the baryonic Tully-Fisher relation given in
Catinella et al (2012) ($ \log V_{max}= 0.237 \log x -0.251; x= M_* + 1.4 M_{HI}$).

\item {\em The dark-matter domination radius} $R_{50}(DM)$ is defined as the radius where the enclosed dark matter
mass exceeds the total baryonic mass. The total mass enclosed within radius R can be estimated directly
from the rotation curve after correction for inclination effects. The inclination of the galaxy is estimated using
its $r$-band  axial ratio $b/a$ as $\cos i = b/a$. 
The total baryonic mass enclosed within radius $R$
is the sum of the stellar and gas mass within this radius. Stellar mass-to-light profiles for each of the galaxies in                   
our sample are  derived by fitting SDSS 5-band optical photometry to stellar population 
synthesis model grids using standard techniques (see for example Salim et al 2005). 
The mass-to-light ratio will depend on the assumed initial mass function (IMF).
For simplicity, we have assumed a fixed Chabrier (2003) form.

Because we only have single-dish measurements of the 21cm and CO(1-0) lines,
the gas mass within radius $R$ must be approximated using assumed profiles.
It has been known for some time that there is a tight relation between the characteristic size
of the HI disk $D1$ and the total HI mass in the disk (Broeils \& Rhee 1997),
\begin {equation} \log M_{HI} =1.96 \log D1 +6.52, \end {equation}
where $D1$ is defined
as the diameter  where the face-on corrected angular-averaged HI column density reaches 1 $M_{\odot}$ pc$^{-2}$ 
(corresponding to $1.25 \times 10^{20}$ atoms cm$^{-2}$). This relation has recently been demonstrated to
extend up to total HI masses of $2 \times 10^{10} M_{\odot}$ and disk diameters of 100 kpc (Wang et al 2013).
Wang et al (2014) show that a simple analytic expression of the form
\begin {equation}
\Sigma_{HI}(r) = \frac{I_1 \exp (-r/r_s)} {1 +I_2 * \exp (-r/r_c)}, 
\end {equation}
where $I_1$, $I_2$, $r_s$ and $r_c$ are free parameters, can provide an excellent description of
the HI density profiles of nearby gas-rich and normal spiral galaxies. Bigiel \& Blitz (2012) show
that total (atomic+molecular) gas profiles are well-characterized by a function of the form
\begin {equation} 
\Sigma_{gas}/\Sigma_{transit} = 2.1 e^{-1.65 r/ R_{25}}
\end {equation}
where $\Sigma_{transit}$ is the radius where the atomic gas surface density equals the molecular
gas surface density, and $R_{25}$ is the optical radius of the galaxy. 

We construct atomic and molecular gas surface density profiles for the galaxies in our sample as
follows. Equation (2) allows us to specify D1, given the observed total HI mass of the
galaxy. Figure 6 of Wang et al (2014) shows that $r_s/R1 = r_s/(0.5 D1)$ lies in the range
0.15 to 0.3. Also based on this data,  we allow $r_c$ to range from 0.05$r_s$ to 0.7$r_s$.
Finally,  we also require that $\int 2 \pi \Sigma_{HI}(r) r dr = M_{HI}$. 
We first generate a family of HI profiles that satisfy these constraints. If the galaxy has
detected molecular gas, then we add in molecular gas at each radius such that equation (4)
is satisfied and we only keep those profiles where the integral over the resulting
molecular gas profile yields a value close to the observed total $H_2$ mass.
Finally we add each acceptable gas profile to the stellar profile and compute  $R_{50}(DM)$
as the value averaged over each $\Sigma_*(r) + \Sigma_{gas}(r)$ combination. In practice, because
stars dominate the baryonic mass budget for the majority of the galaxies in our sample, the uncertainty
in  $R_{50}(DM)$ arising from the variety of acceptable gas profiles is small, typically 10-20 \%.    

\item {\em The spin parameter $\lambda$.} Observational measurements of the specific angular momentum in
galaxies place constraints on the way galaxies acquire their mass and angular momentum. The specific
angular momentum is often parametrized in terms of a dimensionless spin parameter $\lambda$. Here,
we adopt the definition of galaxy spin parameter given in equation (2) of  Dutton \& Van den Bosch  (2012)
\begin {equation} \lambda_{gal} = \frac {(J_{gal}/M_{gal})} {\sqrt{2} R_{vir} V_{vir}}, \end {equation}
where $J_{gal}$ is the total angular momentum of the galaxy, $M_{gal}$ is the total baryonic  mass of the galaxy, and
$R_{vir}$ and $V_{vir}$ are the radius and circular velocity of its dark matter halo.
As discussed by Dutton \& Van den Bosch (2012), if $\lambda_{gal}$ can be measured for a representative set
of galaxies, the angular momentum ratio $R_j$, defined as the ratio between
the galaxy specific angular momentum and the  specific angular momentum of the halo,  
\begin {equation}
R_j= \frac{(J_{gal}/M_{gal})}{(J_{vir}/M_{vir})}= \frac{\lambda_{gal}}{\lambda_{halo}}, 
\end {equation}
can then be constrained.
The distribution of $\lambda_{halo}$ has been extensively studied using N-body simulations (e.g. Bett et al 2007)
and semi-analytic models of disk galaxy formation (e.g. Fu et al 2010, Guo et al 2011) typically {\em assume} that $R_j=1$, because
this leads to good agreement with the observed sizes of galaxies. In this paper, we examine the validity
of this assumption.

Calculating the total angular momentum for each galaxy in our sample, given its rotation curve and baryonic mass density
profile is straightforward:
\begin {equation}
J_{gal} = 2 \pi \int (\Sigma_*(r) +\Sigma_{gas}(r)) V_{rot}(r) r dr, \end {equation}
Accurate estimates of $R_{vir}$ and $V_{vir}$ cannot be made for individual galaxies in our sample. 
Instead, we adopt the parameterized relation between galaxy stellar mass and dark matter halo mass at $z=0$ derived using 
the galaxy abundance-matching technique  
in Moster et al (2013),
\begin {equation} 
M_*/M_{halo}= 2N \left[ (M_{halo}/M_1)^{-\beta} +(M_{halo}/M_1)^\gamma \right]^{-1}, \end {equation} 
with $N=0.035$, $M_1=3 \times 10^{11} M_{\odot}$, $\beta=1.5$ and $\gamma=0.6$. 
The virial radius $R_{vir}$ can then be computed from the relation $M_{halo}= 200 \rho_{crit} (4\pi/3) R^3_{vir}$,
where $\rho_{crit}$ is the critical density of the universe, and $V_{vir}$ is given by $V_{vir}= (G M_{halo}/R_{vir})^{1/2}$.
\end {enumerate}

\section {Results}

\subsection {Inner slope and core radius trends}
In Figure 4, we show plots of the inner slope (with velocity scaled by $V_{max}$ as described in the previous section, and
radius scaled by the half-light radius of the galaxy R$_{50}$), as  functions of a variety of 
different global galaxy parameters. Black points show inner slope values measured for individual galaxies, while
red squares show the running median. We have binned the points along the x-axis such that there are
always 12 galaxies in each bin, and the errorbar on the median value of the slope is calculated by
boot-strap resampling the galaxies in each bin.

\begin{figure*}
\includegraphics[width=125mm]{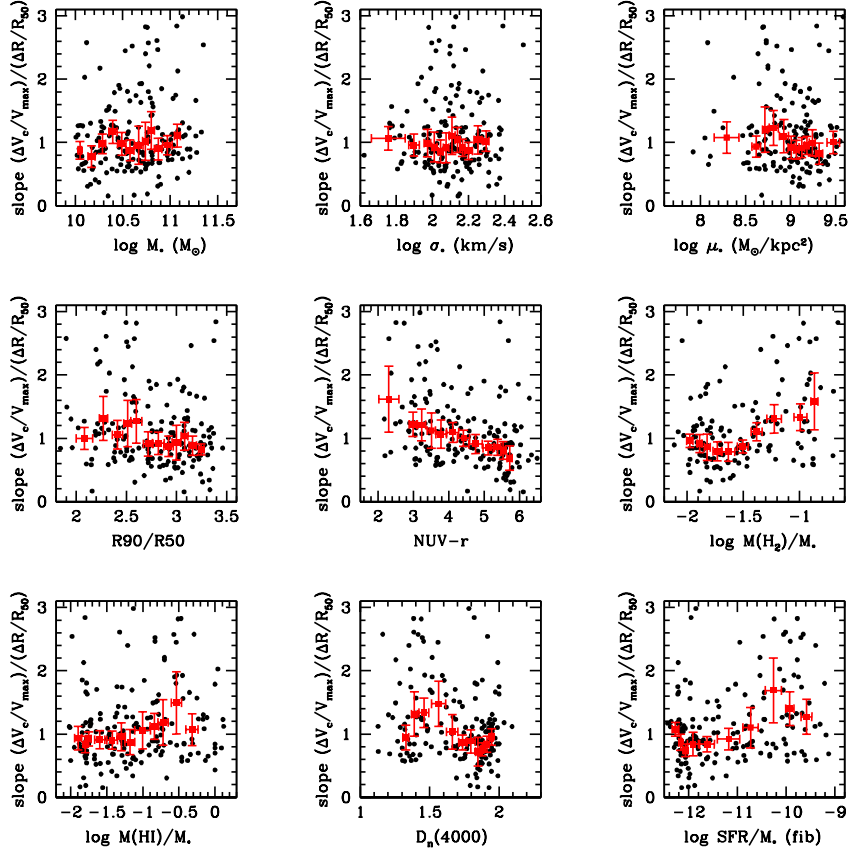}
\caption{ The inner slope of the rotation curve (see text) is plotted as a function of
(from left-to-right and from top-to-bottom)
a)stellar mass $M_*$, b) stellar velocity dispersion measured from the SDSS spectrum, 
c) stellar mass surface density $\mu_*$, d)concentration index, e) NUV-r colour, f)molecular
gas mass fraction, g)atomic gas mass fraction, h)4000 \AA\ break strength, i)specific
star formation measured within the SDSS fibre.  
Black points show inner slope values measured for individual galaxies, while
red squares show the running median. We have binned the points along the x-axis such that there are
always 12 galaxies in each bin, and the errorbar on the median value of the slope in calculated by
boot-strap resampling the galaxies in each bin.
\label{models}}
\end{figure*}

As can be seen from the first two panels, there is no significant trend in inner slope as a function of the stellar mass or stellar
velocity dispersion of the galaxy. The inner slope also does not appear to vary as a function of galaxy structural
parameters such as the stellar surface mass density or the concentration index of the light. The inner slope 
is, however, quite strongly correlated with parameters describing the age of the stellar populations
and the gas content of the galaxy. We note that the detailed behaviour of this trend is different for
each of the 5 parameters we plot. A  roughly continuous trend is seen as a function of the global
NUV-r colour, with the bluest galaxies exhibiting inner slopes that are twice as steep as the
reddest galaxies. For other parameters such as atomic and  molecular gas mass fraction,  specific star
formation rate and 4000 \AA\ break strength, no dependence is seen for the gas-poor/old galaxy population, but
a dependence does set in at higher gas mass fractions and at young ages. The 4000 \AA\ break index D$_n$(4000) and
the specific star formation rate
SFR/$M_*$ are both measured within the 3 arcsecond SDSS fibre aperture and hence probe the stellar populations
in the very central regions of the galaxy. As can be seen in the last two panels, there is again a {\em decrease} in inner slope
for galaxies with the very youngest central stellar populations.     

\begin{figure}
\includegraphics[width=85mm]{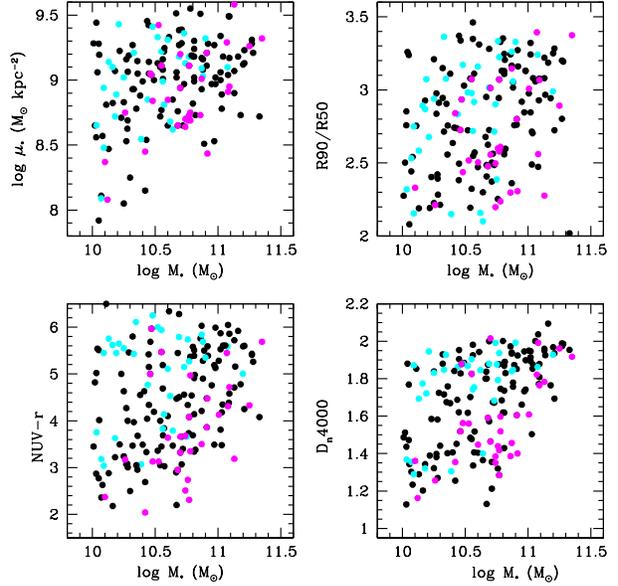}
\caption{ 
Galaxies are plotted in the two-dimensional planes of stellar surface mass density, concentration index,
NUV$-r$ colour and 4000 \AA\ break strength versus stellar mass. Each point is colour-coded according
to the value of the inner slope: cyan points correspond to inner slope values of less than 0.05, black points to
inner slopes in the range 0.05 to 0.2, and magenta points to inner slopes greater than 0.2.
\label{models}}
\end{figure}

In Figure 5, we plot our galaxies in the two-dimensional planes of stellar surface mass density, concentration index,
NUV$-r$ colour and 4000 \AA\ break strength versus stellar mass, and we colour-code each point according
to the value of the inner slope: cyan points correspond to inner slope values of less than 0.05, black points to
inner slopes in the range 0.05 to 0.2, and magenta points to inner slopes greater than 0.2.  
In agreement with the results shown in Figure 4, the galaxies with small and large inner slopes
separate most clearly in NUV-r and D$_n$(4000). However, in this plot we also see that at fixed colour,
stellar surface density or concentration index, galaxies with 
steep inner slopes are more massive than galaxies with flat inner slopes.
This plot helps us to understand  results in the literature (e.g. Persic et al 1996),
which strongly  emphasize the dependence of rotation curve shape on galaxy luminosity or mass. The strong trend with
colour will only be found in samples like ours that include weakly star-forming and early-type galaxies.

A quantity that is closely related to the inner slope is the  radius $r_s$, defined in equation (1). This is the
radius where the rotation curve flattens away from the inner power-law inner region. We note that our definition of
the inner slope of the rotation curve is mainly controlled by variations in the quantity $r_s/R_{50}$,
because $\Delta V_c/ V_{max}$ is generally $\sim 1$. $r_s$ can be regarded as the ``core radius'' of the sum of
the dark matter halo and stellar density profiles. Our results imply that blue, star-forming galaxies
have {\em total mass distributions} with smaller core radii than red, passive galaxies. We note that previous
work (e.g. Kormendy \& Freeman 2004; Spano et al 2008;  Donato et al 2009) has focused on the inferred core radii
dark matter halos after subtracting the contribution from the stars and gas. As we have already discussed, most
of the galaxies in our sample are gravitationally dominated by baryons in their inner regions, so
we will not attempt such a decomposition for the whole sample, only for the subset which are
dark matter dominated within $R_{50}$.

\subsection  {The dark-matter domination radius} 
We are able to follow rotation curve out to large enough
radius for 103 out of the 187 galaxies in our sample to measure the radius $R_{50}(DM)$ where the dark matter 
first contributes more than half the mass interior to that radius. The properties of this subset of galaxies are
shown in Figure 6. Black histograms show distributions of stellar mass, stellar surface mass density,
concentration index, 4000 \AA\ break strength, colour and atomic gas mass fraction for all the galaxies
where we are able to estimate  $R_{50}(DM)$, while red histograms show the distributions of these
properties for galaxies where we do not trace the rotation curve far enough to make an estimate of
this quantity. As can be seen, the lowest mass galaxies with $M_* < 2 \times 10^{10} M_{\odot}$
are missing from the sample with measured  $R_{50}(DM)$, but otherwise the sample  covers a wide
range in colour and structural properties.  

Figure 7 shows how a variety of galaxy parameters correlate with  $R_{50}(DM)$. Once again, we find
weak correlations with galaxy mass and structural parameters and stronger correlations with 
galaxy colour and gas content. The two bottom panels of the figure show the parameters that
correlate best with the dark matter domination radius: the global NUV-r colour and the atomic gas
mass fraction. As can be seen, the majority of galaxies with atomic gas mass fractions greater than 0.1
and NUV-r colours less than 4 become dark matter-dominated within 2$R_{50}$. Galaxies with less than a few
percent atomic gas and NUV-r$> 5$ become dark matter dominated at radii larger than  2$R_{50}$.  
 
\begin{figure}
\includegraphics[width=85mm]{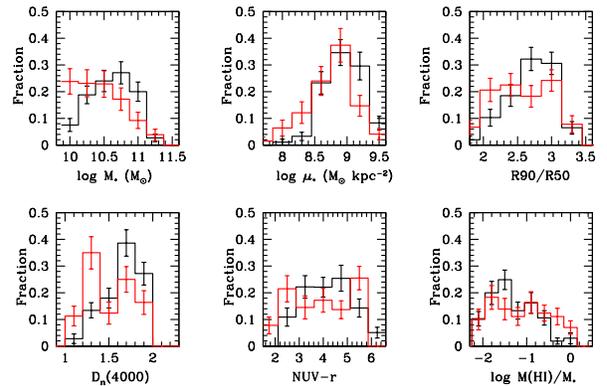}
\caption{ 
Black histograms show the distribution of properties of the subset of 103 galaxies for which we are able to
estimate the dark matter domination radius, while red histograms show the distribution of the
properties of galaxies where the rotation curve does not extend far enough.       
\label{models}}
\end{figure}

\begin{figure}
\includegraphics[width=85mm]{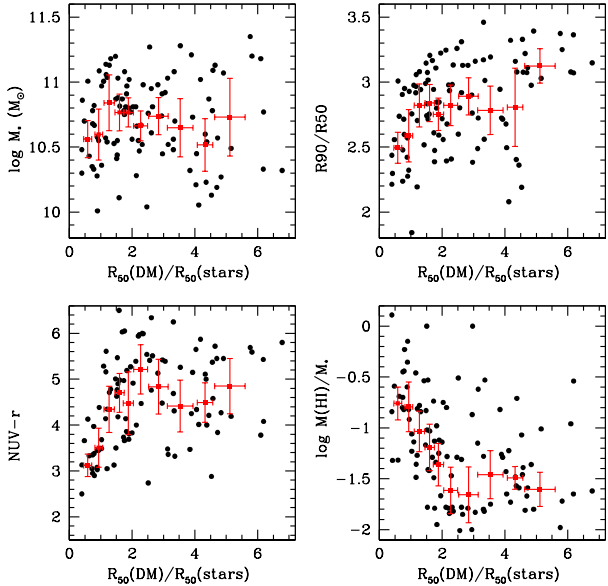}
\caption{ The stellar masses, concentration indices, NUV-r colours and atomic gas mass fractions
are plotted as a function of the dark matter-domination radius  $R_{50}(DM)$.
Black points show parameter values  for individual galaxies, while
red squares show the running median of each parameter at fixed  $R_{50}(DM)$. We have binned the points along the x-axis such that there are
always 12 galaxies in each bin, and the errorbar on the median is calculated by
boot-strap resampling the galaxies in each bin.
\label{models}}
\end{figure}

\subsection {Spin parameter}
In Figure 8, we show plots of the differential (left) and cumulative (right) distributions of  derived spin parameters for 
all 187 galaxies with well-measured rotation curves in our sample. We are able to use the full sample, because
in general, the stellar rotation curves can be accurately determined over most of the optical radius of the
disk. As can be seen, the median value of $\lambda$ is 0.01 and the $\pm 1 \sigma$ range in $\lambda$ extends from  
0.003 to 0.03. The observed median is a factor 3-4 lower than the median value of the spin parameters of  dark matter halo 
measured in N-body simulations of structure formation in a $\Lambda$CDM cosmology 
(e.g. Maccio et al 2007) and the $1 \sigma$ scatter is about twice as large as found in simulations.
Our results are consistent with the conclusions 
presented in  Dutton \& Van den Bosch (2012). These authors inferred  spin parameter distributions indirectly using
ensembles of bulge+disk+halo models that were tuned to fit observational constraints such as the Tully-Fisher relation, the
disk size versus stellar mass relation, and the relation between stellar and dark matter halo mass.      
Based on this analysis, Dutton \& Van den Bosch (2012) inferred  that the average value of $\lambda$
for disk galaxies  was 0.019 and that galaxy spin did not correlate with halo mass.
The big advantage of deriving the spin parameter directly from the observed rotation curve of the galaxy, is that
we are able to examine correlations of $\lambda$ with different galaxy parameters. 

\begin{figure}
\includegraphics[width=85mm]{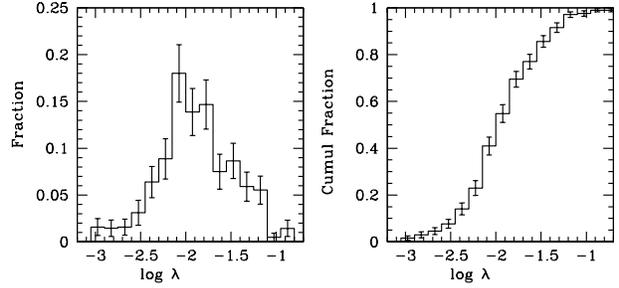}
\caption{ The differential (left) and cumulative (right) distributions of the spin parameters $\lambda$ derived
for the 187 galaxies in our sample with well-measured rotation curves.
\label{models}}
\end{figure}

In Figure 9, we plot $\lambda$ as a function of galaxy stellar mass, concentration index, atomic gas mass fraction and
4000 \AA\ break index strength. Once again, black dots show results for individual galaxies, while the red line shows 
the running median.  The spin parameter correlates with all of the plotted parameters. Low mass, low concentration, gas-rich
galaxies with young stellar populations have larger spin parameters than high mass, high concentration, gas-poor galaxies with
old stellar populations. This is not unexpected since it is believed that the formation of bulge-dominated (early-type) galaxies
involves processes such as interactions or mergers, which cause gas to lose angular momentum (e.g. Barnes \& Hernquist 1996)
Interestingly, galaxies with the lowest masses and highest atomic gas mass fractions have a median spin parameter
of $\sim 0.03-0.04$, which is much more consistent with the predictions of disk formation models where
the gas conserves its angular momentum. In addition, we infer from Figure 9 that the scatter in spin parameter for
galaxies that are similar in stellar mass, structural parameters and stellar population properties is considerably
smaller than the scatter in $\lambda$ for the population as a whole. Larger samples are needed for accurate estimates of the scatter in
$\lambda$ for galaxies with fixed properties.

\begin{figure}
\includegraphics[width=85mm]{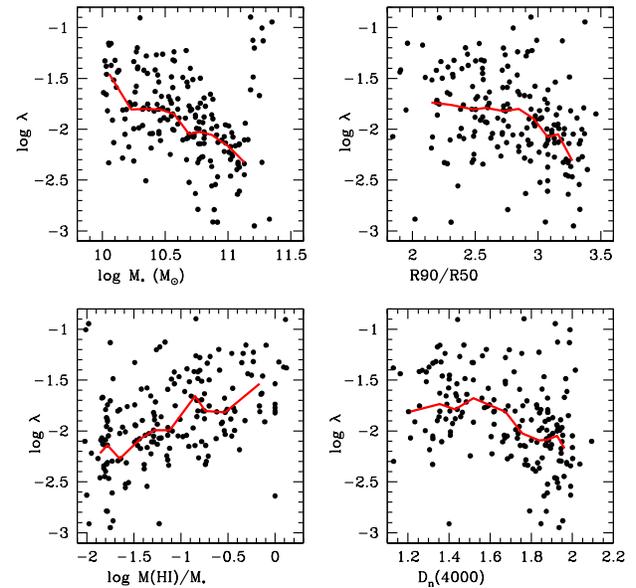}
\caption{ The spin parameter $\lambda$ is plotted as a function of stellar mass, concentration index,  atomic gas mass fraction
and 4000 \AA\ break strength.
Black points show spin parameter values  for individual galaxies, while
the red line shows the running median of the spin as a function of these parameters. 
\label{models}}
\end{figure}

\section {Dark matter halo profile fits}
In two seminal papers, Navarro, Frenk \& White (1996,1997) introduced the concept 
of a ``universal'' profile (hereafter NFW profile) that was able to describe the dark
matter density profiles of halos with masses ranging from those of dwarf galaxy halos to those of rich galaxy clusters,
independent of the  initial density fluctuation spectrum and the values of the cosmological parameters.
The NFW profile has the form
\begin {equation}
\frac {\rho(r)}{\rho_{crit}} = \frac {\delta_c} {(r/r_s) (1+ r/r_s)^2}, \end {equation}
where $r_s = r_{200}/c$ is a characteristic radius and $\rho_{crit} = 3 H^2/8\pi G$ is the critical density
($H$ is the current value of the Hubble constant); $\delta_c$ and $c$ are two dimensionless parameters, commonly
termed the characteristic overdensity and concentration.
The virial mass and radius of the halo are defined through the relation $M_{vir} =200 \rho_{crit} (4\pi/3)R_{vir}^3$, and
$\delta_c$ and $c$ are then linked by the requirement that the mean density within $R_{vir}$ is $200 \rho_{crit}$.  

Rotation curve fitting is one important way to test whether the NFW profile can fit the dark matter halos of real
galaxies. There are many papers in the literature with seemingly contradictory results, but the most
recent consensus appears to be that the NFW profile is a reasonably good fit for luminous galaxies (De Blok et al 2008;
Martinsson et al 2013), but is a poorer fit for low surface brightness dwarf systems, where the dark matter
strongly dominates over the baryonic mass (e.g. Swaters et al 2003). One problem with the interpretation of
the results for dwarf galaxies is that because the potential well depth is shallow, feedback effects from
bursts of star formation during the formation of the dwarf can in principle affect the 
dark matter density profile (e.g. Pontzen \& Governato 2012).

In this section, we perform NFW halo fits to a subset of 12 galaxies in our sample where the dark matter domination
radius  $R_{50}(DM) < 0.7 R_{50}$. In the Milky Way, dark matter dominates over the baryons at radii larger than
$\sim 10$ kpc, i.e. at around 3 R$_{50}$. The galaxies that we analyse are thus much more dark-matter
dominated than our own Galaxy and as shown in Figure  7, they also tend to have young stellar
populations and to be more gas-rich. A detailed analysis of one  such a galaxy in our sample, UGC8802 (GASS 35981), was presented in Moran et al (2010).     
In spite of its high atomic gas content ($2.1\times10^{10} M_{\odot}$), the  star formation surface density in this galaxy is low
($\Sigma_{SFR}= 0.003 M_{\odot}$ yr$^{-1}$ kpc$^{-2}$) and spread evenly across the galaxy. A sharp  drop in metallicity in the outer disk 
was found in the galaxy, which was later shown to be a common feature of many of the  HI-rich galaxies in the sample
(Moran et al 2012).

In Figures 10, 11, and 12, we present the NFW halo fits for our 12 selected dark matter-dominated galaxies. The measured values of the the rotation
velocity are plotted as black circles. The dotted red curves  show the contribution from the observed baryons (stars+gas), while the solid
red curves show the best fit rotation curve derived from adding the contribution from a NFW halo to the baryons. \footnote {We have not
attempted to correct for ``adiabatic contraction'', the gravitational effect of the central baryonic contribution on the dark matter.}  
The parameters of the best-fit NFW model are indicated in the bottom-right corner of each panel. 

As can be seen, the rotation curves of these 12 galaxies are regular for the most part, and 
the NFW halo+observed baryon models provide reasonably good fits to all the galaxies in our sample.
There are no cases where the model fails catastrophically and in general the inner power-law region of the rotation
curve is well-reproduced. We note that unlike many other studies, we allow no freedom in the assumed mass-to-light
ratios for the baryonic component; as described in Section 2,  these are derived from fits to the $u,g,r,i,z$
spectral energy distributions, assuming a universal Chabrier (2003) initial mass function.
As an additional check, we compare the rotational velocity derived from the  width of the 21cm line (see Catinella et al 2012 for details)
with the velocity width predicted from weighting the rotation curve by our assumed HI profiles (see section 2). The  value measured from
the Arecibo spectrum is plotted as a black star in each panel, while the value inferred from the rotation curve
is plotted as a red star. As can be seen, the two estimates generally agree to within 10-20\% and confirm that our estimates
of $V_{vir}$ are reasonably accurate. All the galaxies in our sample are found to reside in quite massive dark matter halos:
the smallest (GASS 3591) has a virial velocity of 185 km/s, corresponding to a virial mass of $2 \times 10^{12} M_{\odot}$.
The average value of the halo concentration index $c$ is found to be 9.6 and the 10th-90th percentile range
is from 6 to 14. This is in excellent agreement with predictions from $\Lambda$CDM  cosmological simulations
(see Figure 3 of Maccio et al (2007)).

In summary, we conclude that NFW halos provide a good fit to the inner halos of the most dark matter-dominated galaxies in our sample.

\begin{figure}
\includegraphics[width=85mm]{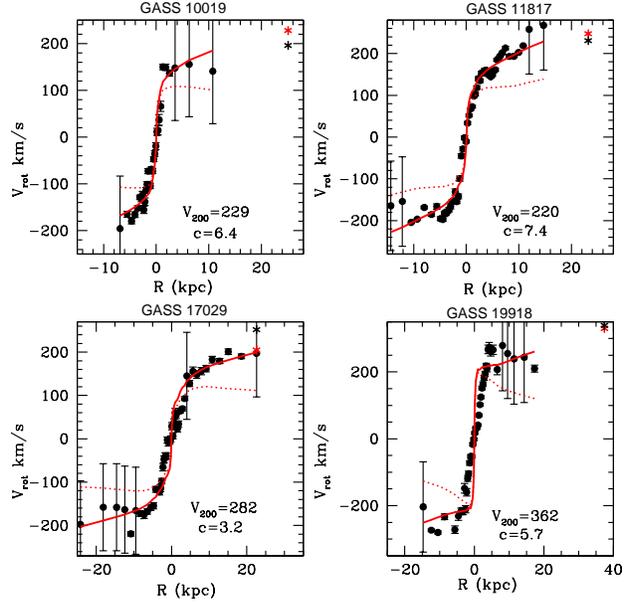}
\caption{ NFW halo fits to the rotation curves of four dark-matter dominated galaxies.                                                                
The measured values of the rotation
velocity are plotted as black circles. The dotted red curves  show the contribution from the observed baryons (stars+gas), while the solid
red curves show the best fit rotation curve derived from adding the contribution from a NFW halo to the baryons.
The HI-profile weighted value of $V_{rot}$  measured from
the Arecibo spectrum is plotted as a black star in each panel, while the value inferred from the rotation curve
is plotted as a red star.
\label{models}}
\end{figure}

\begin{figure}
\includegraphics[width=85mm]{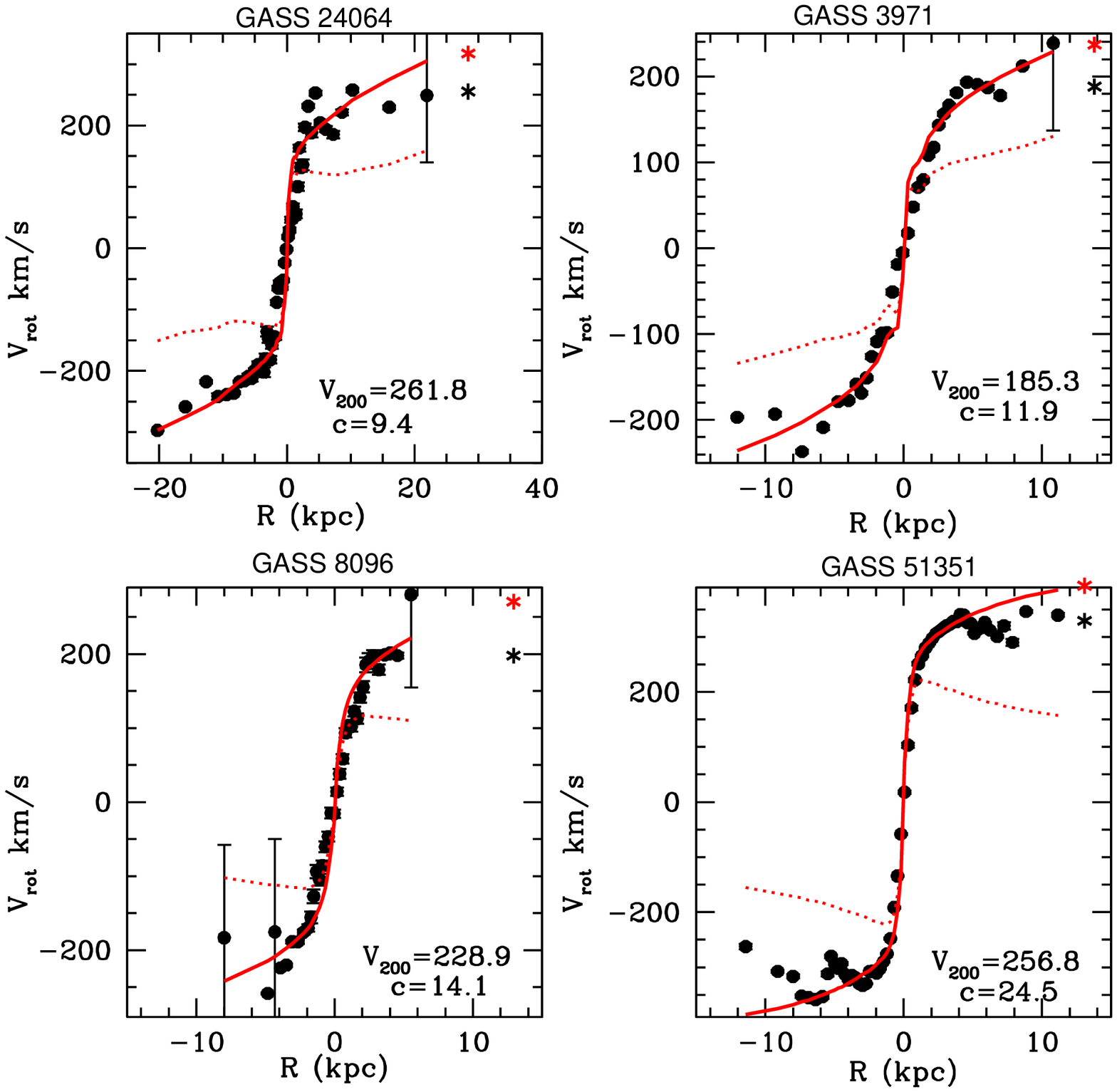}
\caption{Same as Figure 10, but for 4 more dark-matter dominated galaxies.                                                                                                 
\label{models}}
\end{figure}

\begin{figure}
\includegraphics[width=85mm]{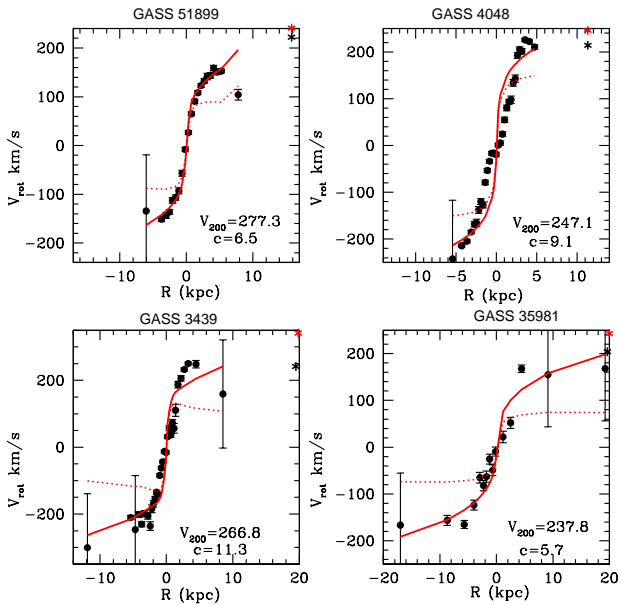}
\caption{Same as Figure 11, but for 4 more dark-matter dominated galaxies.                                                                                                                           
\label{models}}
\end{figure}

\section {Summary and discussion}
The main empirical results of our rotation curve study are the following:
\begin {itemize}
\item The inner slopes  of the rotation curves correlate  more  strongly with measures of stellar population age
than with galaxy mass or structural parameters. At fixed stellar mass, more actively star-forming  galaxies
have steeper inner slopes  than less actively star-forming galaxies. 
Another way to phrase this result is that the ratio of the radius $r_s$, where the rotation curve
transitions away a power-law, to the half-light radius of the galaxy, is smaller in star-forming
galaxies than in passive galaxies.             

\item Galaxies with higher atomic gas mass fractions tend to be more dark-matter dominated than galaxies
with low HI content. 

\item The distribution of spin parameters of the galaxies in our sample shows that the average
value of the spin parameter of the galaxies in our sample is $\sim 0.01$, which
implies that the  angular momentum ratio $R_j$, defined as the ratio between the galaxy specific angular
momentum and the total specific angular momentum of its dark matter halo, is less than 1.
However, $R_j$ is found to correlate strongly with galaxy mass, structure and gas content.
The lowest mass, disk-dominated  galaxies with atomic  gas mass  fractions greater than 20\%
have median values of  $R_j $ of around 1. The highest mass, gas-poor, bulge-dominated galaxies 
have median values of $R_j$ of 0.1-0.2, i.e. such galaxies have substantially lower specific angular momenta 
than their host halos.     

\item The dark matter halo density profiles of the  most dark-matter dominated galaxies in our sample 
are well fit by NFW models. The  average concentration parameter is $\sim 10$, in good agreement
with the predictions of N-body simulations of structure formation in a $\Lambda$CDM Universe. 
\end {itemize}

One might first ask whether all of these trends are qualitatively consistent with  a picture in which
galaxies form by gas cooling and condensation within dark matter halos in a canonical $\Lambda$CDM cosmology.
In such a picture, the reason why $R_j$ would be less than 1 in some galaxies could
be because gas had lost angular momentum and    
had been channelled inwards in these systems.  
This inward flow of gas would  
result in  
a build-up of higher density gas in the inner regions of the disk, which  would then rapidly cool to
low temperatures and  form bulge stars.  Gas that cools and accretes from the
surrounding halo is  used up more
quickly  if gas inflows in the disk are common.  Frequent inflows would
thus be expected to lead to a galaxy that is passive and bulge-dominated.   
This is a possible explanation of  our finding that the spin parameter is smaller in
bulge-dominated galaxies with low gas content and old stellar populations.

Gas inflows can arise in two ways: 1) when the disk is perturbed by a merger or interaction with
another galaxy, 2) when the disk becomes unstable and forms a bar. The Toomre (1964) disk stability
parameter $Q$ is larger in galaxies with higher epicyclic frequencies and larger shear (i.e. larger
$d V_{rot}/dR$). We find that the inner slope of the rotation curve is large in galaxies that are
still forming stars, which is consistent with the idea that star formation can be sustained for
longer periods if the disk is more stable.

The next step is to test these ideas quantitatively by comparing our results with
N-body + hydrodynamical simulations of disk galaxy formation in a $\Lambda$CDM Universe in which
the cooling of gas in dark matter halos  and its subsequent dynamical evolution in disks is
treated in a physically self-consistent framework. Well-resolved, dynamically consistent simulations                        
rather than simplified analytic models of disk formation in a hierarchical Universe are also required
to predict the effects
of feedback processes from supernova and accreting black holes on the inner mass
distributions of galaxies of different types and star formation rates.  
They will also be very helpful in developing methodology for correcting for 
asymmetric drift.

There are also hints of more complex dependences of rotation curve shape on star formation history. 
In Figure 4,  we found that galaxies with the very youngest central stellar populations
have rotation curves with shallower inner slopes. This hints that supernovae feedback processes
occurring during strong central starbursts may be able to alter the
inner mass distribution of galaxies. It has been proposed that such mechanisms may be key for
resolving the observed rotation curves to dwarf galaxies with the dark matter
halo profile predictions of the Cold Dark Matter model.
Dynamical studies of larger galaxy samples   
observed using integral field spectrographs spanning a wider range of masses and redshifts will shed more light
this issue.

%===================================
\section*{Acknowledgments}
We thank Simon White for helpful discussions.

%===================================


\begin{thebibliography}{}

\bibitem[\protect\citeauthoryear{Barnes 
\& Hernquist}{1996}]{1996ApJ...471..115B} Barnes J.~E., Hernquist L., 1996, ApJ, 471, 115 

\bibitem[\protect\citeauthoryear{Bett et al.}{2007}]{2007MNRAS.376..215B} 
Bett P., Eke V., Frenk C.~S., Jenkins A., Helly J., Navarro J., 2007, 
MNRAS, 376, 215 

\bibitem[\protect\citeauthoryear{Bigiel 
\& Blitz}{2012}]{2012ApJ...756..183B} Bigiel F., Blitz L., 2012, ApJ, 756, 183 

\bibitem[\protect\citeauthoryear{B{\"o}hm et 
al.}{2004}]{2004A&A...420...97B} B{\"o}hm A., et al., 2004, A\&A, 420, 97

\bibitem[\protect\citeauthoryear{Broeils 
\& Rhee}{1997}]{1997A&A...324..877B} Broeils A.~H., Rhee M.-H., 1997, A\&A, 324, 877 

\bibitem[\protect\citeauthoryear{Bruzual 
\& Charlot}{2003}]{2003MNRAS.344.1000B} Bruzual G., Charlot S., 2003, MNRAS, 344, 1000 

\bibitem[\protect\citeauthoryear{Catinella et
al.}{2010}]{2010MNRAS.403..683C} Catinella B., et al., 2010, MNRAS, 403,
683

\bibitem[\protect\citeauthoryear{Catinella et 
al.}{2012}]{2012MNRAS.420.1959C} Catinella B., et al., 2012, MNRAS, 420, 
1959 

\bibitem[\protect\citeauthoryear{Chabrier}{2003}]{2003PASP..115..763C} 
Chabrier G., 2003, PASP, 115, 763 

\bibitem[\protect\citeauthoryear{Combes}{2002}]{2002NewAR..46..755C} Combes 
F., 2002, NewAR, 46, 755 

\bibitem[\protect\citeauthoryear{de Blok et 
al.}{2008}]{2008AJ....136.2648D} de Blok W.~J.~G., Walter F., Brinks E., 
Trachternach C., Oh S.-H., Kennicutt R.~C., Jr., 2008, AJ, 136, 2648 

\bibitem[\protect\citeauthoryear{Donato et al.}{2009}]{2009MNRAS.397.1169D} 
Donato F., et al., 2009, MNRAS, 397, 1169 

\bibitem[\protect\citeauthoryear{Dutton 
\& van den Bosch}{2012}]{2012MNRAS.421..608D} Dutton A.~A., van den Bosch F.~C., 2012, MNRAS, 421, 608 

\bibitem[\protect\citeauthoryear{Fu et al.}{2010}]{2010MNRAS.409..515F} Fu 
J., Guo Q., Kauffmann G., Krumholz M.~R., 2010, MNRAS, 409, 515 

\bibitem[\protect\citeauthoryear{Guo et al.}{2011}]{2011MNRAS.413..101G} 
Guo Q., et al., 2011, MNRAS, 413, 101 

\bibitem[\protect\citeauthoryear{Kormendy 
\& Freeman}{2004}]{2004IAUS..220..377K} Kormendy J., Freeman K.~C., 2004, IAUS, 220, 377 

\bibitem[\protect\citeauthoryear{Macci{\`o} et 
al.}{2007}]{2007MNRAS.378...55M} Macci{\`o} A.~V., Dutton A.~A., van den 
Bosch F.~C., Moore B., Potter D., Stadel J., 2007, MNRAS, 378, 55 

\bibitem[\protect\citeauthoryear{Martin}{2005}]{2005ApJ...621..227M} Martin 
C.~L., 2005, ApJ, 621, 227 

\bibitem[\protect\citeauthoryear{Martinsson et 
al.}{2013a}]{2013A&A...557A.130M} Martinsson T.~P.~K., Verheijen M.~A.~W., Westfall K.~B., Bershady M.~A., Schechtman-Rook A., Andersen D.~R., Swaters R.~A., 2013, A\&A, 557, AA130 

\bibitem[\protect\citeauthoryear{Martinsson et 
al.}{2013b}]{2013A&A...557A.131M} Martinsson T.~P.~K., Verheijen M.~A.~W., Westfall K.~B., Bershady M.~A., Andersen D.~R., Swaters R.~A., 2013, A\&A, 557, AA131 

\bibitem[\protect\citeauthoryear{Moran et al.}{2007}]{2007ApJ...659.1138M} 
Moran S.~M., Miller N., Treu T., Ellis R.~S., Smith G.~P., 2007, ApJ, 659, 
1138 

\bibitem[\protect\citeauthoryear{Moran et al.}{2010}]{2010ApJ...720.1126M} 
Moran S.~M., et al., 2010, ApJ, 720, 1126 

\bibitem[\protect\citeauthoryear{Moran et al.}{2012}]{2012ApJ...745...66M} 
Moran S.~M., et al., 2012, ApJ, 745, 66 

\bibitem[\protect\citeauthoryear{Moster, Naab, 
\& White}{2013}]{2013MNRAS.428.3121M} Moster B.~P., Naab T., White S.~D.~M., 2013, MNRAS, 428, 3121 

\bibitem[\protect\citeauthoryear{Navarro, Frenk, 
\& White}{1996}]{1996ApJ...462..563N} Navarro J.~F., Frenk C.~S., White S.~D.~M., 1996, ApJ, 462, 563 

\bibitem[\protect\citeauthoryear{Navarro, Frenk, 
\& White}{1997}]{1997ApJ...490..493N} Navarro J.~F., Frenk C.~S., White S.~D.~M., 1997, ApJ, 490, 493 

\bibitem[\protect\citeauthoryear{Noordermeer et 
al.}{2007}]{2007MNRAS.376.1513N} Noordermeer E., van der Hulst J.~M., 
Sancisi R., Swaters R.~S., van Albada T.~S., 2007, MNRAS, 376, 1513 

\bibitem[\protect\citeauthoryear{Persic, Salucci, 
\& Stel}{1996}]{1996MNRAS.281...27P} Persic M., Salucci P., Stel F., 1996, MNRAS, 281, 27 

\bibitem[\protect\citeauthoryear{Pontzen 
\& Governato}{2012}]{2012MNRAS.421.3464P} Pontzen A., Governato F., 2012, MNRAS, 421, 3464 

\bibitem[\protect\citeauthoryear{Saintonge et 
al.}{2011}]{2011MNRAS.415...32S} Saintonge A., et al., 2011, MNRAS, 415, 32 

\bibitem[\protect\citeauthoryear{Salim et al.}{2005}]{2005ApJ...619L..39S} 
Salim S., et al., 2005, ApJ, 619, L39 

\bibitem[\protect\citeauthoryear{Sofue et al.}{1999}]{1999ApJ...523..136S} 
Sofue Y., Tutui Y., Honma M., Tomita A., Takamiya T., Koda J., Takeda Y., 
1999, ApJ, 523, 136 

\bibitem[\protect\citeauthoryear{Sofue 
\& Rubin}{2001}]{2001ARA&A..39..137S} Sofue Y., Rubin V., 2001, ARA\&A, 39, 137 

\bibitem[\protect\citeauthoryear{Spano et al.}{2008}]{2008MNRAS.383..297S} 
Spano M., Marcelin M., Amram P., Carignan C., Epinat B., Hernandez O., 
2008, MNRAS, 383, 297 

\bibitem[\protect\citeauthoryear{Swaters et 
al.}{2003}]{2003ApJ...583..732S} Swaters R.~A., Madore B.~F., van den Bosch 
F.~C., Balcells M., 2003, ApJ, 583, 732 

\bibitem[\protect\citeauthoryear{Toomre}{1964}]{1964ApJ...139.1217T} Toomre 
A., 1964, ApJ, 139, 1217 

\bibitem[\protect\citeauthoryear{Wang et al.}{2013}]{2013MNRAS.433..270W} 
Wang J., et al., 2013, MNRAS, 433, 270

\bibitem[\protect\citeauthoryear{Wang et al.}{2014}]{2014MNRAS.441.2159W} 
Wang J., et al., 2014, MNRAS, 441, 2159 

\bibitem[\protect\citeauthoryear{Weinmann et 
al.}{2010}]{2010MNRAS.406.2249W} Weinmann S.~M., Kauffmann G., von der 
Linden A., De Lucia G., 2010, MNRAS, 406, 2249 

\bibitem[\protect\citeauthoryear{York et al.}{2000}]{2000AJ....120.1579Y} 
York D.~G., et al., 2000, AJ, 120, 1579 

\end{thebibliography}
\end{document}